\documentclass[conference]{IEEEtran}
\IEEEoverridecommandlockouts
\usepackage{url}
\usepackage{graphicx,epsfig}
\usepackage{subfig}
\usepackage{graphics,dcolumn,bm,epic, eepic,fleqn,float}
\usepackage{amssymb,amsmath,multirow,rotate,color}
\usepackage{epstopdf}
\usepackage{soul}
\usepackage{cleveref}

\usepackage{color}
\usepackage[]{algorithm2e}
\usepackage{booktabs}
\usepackage[T1]{fontenc}
\usepackage{amsmath}

\newcommand\fref[1]{Fig.~\ref{#1}}
\newcommand\tref[1]{Table~\ref{#1}}

\let\oldenumerate\enumerate
\renewcommand{\enumerate}{
  \oldenumerate
  \setlength{\itemsep}{1pt}
  \setlength{\parskip}{0pt}
  \setlength{\parsep}{0pt}
}

\let\olditemize\itemize
\renewcommand{\itemize}{
  \olditemize
  \setlength{\itemsep}{1pt}
  \setlength{\parskip}{0pt}
  \setlength{\parsep}{0pt}
}
\IEEEoverridecommandlockouts

\def\BibTeX{{\rm B\kern-.05em{\sc i\kern-.025em b}\kern-.08em
    T\kern-.1667em\lower.7ex\hbox{E}\kern-.125emX}}
\begin{document}

\title{Evaluation of Availability of Initial-segments of Video Files in Device-to-Device (D2D) Network*\\
}
\author{\IEEEauthorblockN{1\textsuperscript{} Nasreen Anjum}
\IEEEauthorblockA{\textit{ Centre for Telecommunications Research} \\
\textit{King's College London }\\
London, UK \\
nasreen.anjum@kcl.ac.uk}
\and
\IEEEauthorblockN{2\textsuperscript{} Mohammd Shikh Bahaei (Senior Member, IEEE)}
\IEEEauthorblockA{\textit{Centre for Telecommunications Research} \\
\textit{King's College London.}\\
London, UK \\
m.sbahaei@kcl.ac.uk}
}

\maketitle

\begin{abstract}

We propose a new caching concept in which wireless nodes in Device-to-Device (D2D) network can cache the initial-segments of the popular video files and may share to users in their proximity upon request. We term this approach as Bootstrapping-D2D (B-D2D) system. Our findings suggest that caching only initial-segments creates a large pool of popular files, thus improving the probability of availability of initial-segments within the vicinity of the requesting users. Our proposed B-D2D system is based on a popular clustering algorithm called density-based spatial clustering of applications with noise (DBSCAN), that groups the users into dense and arbitrary shape clusters based on two predefined thresholds. We then formulate an optimization problem to maximize the probability of availability of initial-segments within the transmission range of requesting wireless devices. The simulation results demonstrate that, caching the initial video segments could substantially improve the availability of segments within the proximity of users without requiring any extra hardware infrastructure cost, and D2D services could be utilized to deliver the video contents with lower startup delays in cellular networks.
\end{abstract}


\section{Introduction}

Driven by clients' expanding appetite for videos, present period witnesses a gigantic growth in video data traffic over the wireless cellular networks (WSNs). The successful deployment of high-speed fourth generation (4G) technology of mobile networks, multiplication in the smart hand-held devices~\cite{CiscoRep2015}, reasonable boundless data plans~\cite{karamshuk2015factors}, and an assortment of dynamic and interactive services offered by the cellular networks, have resulted in exponential raise in video streaming services being viewed by customers. It is forecasted that mobile video traffic will increase up-to an 11-times by the end of year 2020~\cite{Ciscowhitepaper}.
The aforementioned figure has created opportunities for wireless service providers (WSPs) to maximize their customer size and satisfaction ratio, that may ultimately boost their business revenue. However, in order to achieve this desired goal, WSPs need to keep their audience engaged by providing high quality services such as \emph{lower startup delays}. For instance, a study conducted to analyze the effect of video stream quality on customer behavior reported that longer startup delay is intolerable and compel the users to leave the video unwatched at once~\cite{whyvideo}. As a consequence of this users may abandon the subscribed services. 

Video streaming with guaranteed QoS is one of the central 5G's vision~\cite{panwar2016survey}. Therefore, having a fast startup time for video streaming applications is critical for current and future generation cellular networks. In commercially deployed peer-assisted content delivery networks (PA-CDNs) such as Kankan and LiveSky, QoS in terms of lower startup delay is achieved by bootstrapping video streaming with initial-segments\footnote{In this paper, for the brevity of readers we use the term initial-segments to represent the group of initial frames of a video. For example, initial-segment containing 10sec video clip. The video file is sliced to small segments before deliver to end-user, and segment is manipulated as a minimum unit for caching and delivering the video.} downloaded from the geographically deployed nearest edge server nodes~\cite{anjum2017survey}. Unfortunately, on the other hand,  due to scarcity of radio resources~\cite{shadmand2010multi}~\cite{saki2015cross}~\cite{kobravi2007cross}, and inherent adverse features such as channel fading, interference, mobility and end-to-end delay~\cite{shadmand2009tcp}~\cite{bobarshad2009m}~\cite{bobarshad2012analytical}, currently deployed wireless cellular infrastructure (3G/4G) is failing to meet customers' expectations, and video one demand (VoD) streaming is experiencing unpredictable long buffering delays. For instance, one prior study has reported several hundreds of milliseconds delay, when users download video stream through the BS in a cellular network~\cite{astely2013lte}.

Recently, D2D communication has become an attractive distributed computing paradigm, following the popularity and standardization of unlicensed band D2D protocols in LTE Rel.12~\cite{astely2013lte}. D2D technology enables direct communication between devices located in close proximity without requiring any assistance from mobile core architecture. Some initial studies have investigated the impact of video distribution via D2D on cellular network and suggested to employ the D2D technology to assist BS in distributing the multimedia content. For instance, authors in~\cite{golrezaei2014base} propose to cache and share the viral videos to the proximity users via D2D communication link over the unlicensed frequency band. The simulation results shows that their proposal for video dissemination through D2D communication has the capacity to improve the video throughput by one or two order of magnitude. Another study~\cite{zhang2015social} proposed an algorithm to distribute the Online Social Network's (OSN) multimedia content through D2D communication by exploiting the characteristics of social networks. The simulation results reported an increase in data rates of the system. In principle, exploiting direct communication between nearby devices and their cache capacity, D2D technology promises to improve spectrum utilization, video throughput, energy efficiency, and simultaneously a better user experience ~\cite{golrezaei2014base}~\cite{wang2014cache}~\cite{zhang2015social}~\cite{naslcheraghi2017fd}.
\subsection{Problem Statement}
Even with the benefits the D2D technology promises to deliver, there are some obstacles which need to be considered while adopting it widely. For instance, 
\begin{itemize}
\item Firstly, mobile devices are highly transient in nature.
\item Secondly, although storage space on wireless devices is increasing, yet they have heterogeneous and finite cache capacities. \item Thirdly, the size and popularity of the video files are also highly dynamic and mobile devices sometimes can cache and share only fraction of the video file. 
\end{itemize}

In such cases provision of good quality video streaming through D2D communication cannot be guaranteed~\cite{wang2014cache}. 
Therefore, due to aforementioned constraints and the potential benefits of D2D communication reported in ~\cite{golrezaei2014base}~\cite{wang2014cache}~\cite{zhang2015social}, such as improvement in video throughput and data transfer rates, we propose to cache and deliver the initial-segments of popular video files via D2D communication in a cellular network. 


\subsection{Contributions}

This article aims to obtain the probability of availability of initial-segments within the vicinity of the requesting device. We propose a D2D network, in which the wireless users can cache and share initial-segments of the popular video files to their proximal users via a D2D communication link. To the best of our knowledge, this is the first article focusing on investigating the availability of initial-segments in D2D scenario. Our main contributions are as follows: 

\begin{itemize}
\item First, we introduce the concept of the Bootstrapping -D2D system (B-D2D)~\footnote{Bootstrapping term is used in typical P2P and PA-CDN networks to represent nodes that provide initial-segments to requesting device to achieve lower startup delay~\cite{anjum2017survey}.}, that uses the mobile's cache to store and deliver the initial-segments of the popular video files with their proximal users.
\item Second, to provide the analysis of our B-D2D concept, we propose a system based on a density based clustering algorithm called DBSCAN~\cite{ester1996density}. DBSCAN is the most well-known clustering algorithm that groups the users into meaningful and arbitrary shaped clusters based on two predefined thresholds such as "radius" and "minimum number of neighbors" with in a particular "radius". 
\item Third, we formulate the optimization problem that maximizes the probability of availability of initial-segments within the proximity of requesting users. 
\end{itemize}


Our simulation results suggest that, the cache of mobile devices can store the initial-segments of multiple video files subject to availability of storage space and creates a large pool of cache of initial-segments. Thus, have the potential to dramatically increase the probability of finding the initial-segments of desired video contents in close proximity. Since direct communication between nearby devices and their cache capacity provide better video throughput and simultaneously a better user experience ~\cite{golrezaei2014base}~\cite{wang2014cache}~\cite{zhang2015social}, B-D2D approach may reduce the startup delay when a user downloads the initial-segments from the nearby device.  



The remainder of this paper is organized as follows. In section II, we introduce our proposed system along with an analysis, simulation results are presented in section 3, and section IV concludes the paper.

\begin{table}[t!]
\caption{Parameters of the Bootstrapping-D2D system}
\small{
\centering
\begin{tabular}{l|p{6.5cm}}
\toprule
Variable & Description\\
\midrule
$\eta$ &Number of $BSNs$ \\
$\omega$ &Number of base stations \\
$\lambda_{\eta}$ & Density distribution of BSNs\\
$\lambda_{\omega}$ & Density distribution of base stations\\
$\beta$ &Zipf exponent \\
$L$ &Number of files\\
$P^{(Y)}_{F}$ &Popularity distribution of L video files\\
$\epsilon_{max}$ &Maximum transmission range to join a cluster\\
$MinBSN$ & Minimum number of $BSNs$ required to form a cluster\\
$P^{(i)}_{U}$ &Probability of unavailability of initial-segments\\
$C$ & Cache capacity of a $BSN$\\
$P^{(i)}_{r}$ &Request probability of initial-segments\\
$Q_{i}$ & Caching probability of initial-segments\\
$P^{(i)}_{A}$ & Availability probability of initial-segments\\
$S_i$ & Size of initial-segment of content $i$\\
\bottomrule
\end{tabular}
\label{tbl:params}}
\end{table}

\section{Introduction and Analysis of B-D2D System}

In this section, firstly, we introduce the concept of B-D2D video streaming system. Secondly, we briefly review the principles and important characteristics of DBSCAN. Thirdly, we use our proposed approach to formulate the optimization problem to investigate the probability of availability of initial-segments in close proximity of requesting users. The important parameters we used in our proposed system are summarized in \tref{tbl:params}.
\subsection{ Video Streaming and Caching System of B-D2D Approach}
The Boot-Strapping-Node ($BSN$) in our system is a mobile device that can cache the initial-segments of popular video files and may share upon request using D2D opportunities. We term this act of sharing initial video segments from a D2D as Bootstrapping-D2D (B-D2D). We assume that, video file in B-D2D system is composed of a number of small and equal size segments. Each segment contains a few second frames and frame is a group of still pictures (GOP). The first segment which contains few seconds of initial video clip, we called it an initial-segment and is the basic unit of operation. A $BSN$ in our video streaming system maintains a playback buffer that stores all received segments from the network or neighboring $BSNs$. The received segments from different sources are assembled in the playback buffer in playback order. The media player of mobile user displays the content from this buffer. We also assume that all video files are encoded using a constant bit-rate (CBR) technique.

We consider the popularity-based random caching policy in which each node always caches the initial-segments of accessed video file in its local storage according to some popularity distribution. Therefore, B-D2D network may contains the redundant copies of the same video file. The $BSN$ may get and play desired videos from their local storage with zero startup delay through self-request. Once the requesting device has cached the initial-segment and started to play the video, BS will push the remaining segments to the buffer of requesting device for service continuity. We emphasize here that, discussion on service continuity, and other aspects related to quality of service are out of scope of this paper. Moreover, our work in this paper is not a complete caching algorithm, but an investigation, how  caching only initial-segments could improve the probability of availability of initial-segments within the proximity of requesting devices. 


Many studies have reported skewed distribution of users' interest toward small fraction of top ranked content~\cite{li2011measurement}. We prefer to use the Zipf distribution model which has been used in many studies for the analysis of  D2D caching~\cite{golrezaei2014base}~\cite{wang2014cache}. We assume that every user from cluster independently and randomly requests a file from the library of size $L$, and the popularity of files can be expressed as:
\begin{align}
 P^{(Y)}_{F}= \frac{\frac{1}{Y^\beta}}{\sum_{\ell=1}^{L} \frac{1}{\ell^\beta}}, 1 \leq Y \leq L\
\end{align}
where $\beta$ is a value of the exponent characterizing the distribution, $L$ is the number of files in library and $Y$ be their rank.


\subsection{Motivation and Principles for Using the DBSCAN Algorithm} 

\textbf{Motivation:} The video streaming analysis in D2D system typically involves classifying the mobile devices into meaningful and useful clusters based on some related information. We propose to use the DBSCAN to cluster the users in D2D system for many reasons.


Some of the early studies perform the analysis of D2D system by distributing users uniformly in square, equal size~\cite{golrezaei2014base} and spherical shape clusters~\cite{kang2014mobile}. Since wireless users are mobile in nature, any restrictions we impose to the shape of clusters could put strong limitations for such applications, so spherical-shape clustering algorithms (K-mean ~\cite{ahmad2007k}) and square-shape-clustering algorithms (grid-based clustering ~\cite{celebi2015partitional}) could not be generally applied. DBSCAN is particularly well suited for mobile users such as large group of mobile users  moving from one cell to another in the wireless network. In such case, mobility pattern of moving objects would form a irregular-shape cluster (see ~\fref{fig:systemmodel}). DBSCAN has the ability to discover and construct the non-spherical and non-square arbitrary shape clusters.  Moreover, contrarily to spherical shape clustering algorithms, DBSCAN can also easily differentiate between cluster of points and outliers\footnote{An outlier is a data point or wireless node that lies outside any cluster.}, and also with lower complexity (O(nlogn) vs O($n^2$)). 
The key idea of DBSCAN is that, for each mobile device in some cluster has to contain minimum number of neighbors within a given radius $\epsilon$. For that reason, this algorithm is popular for its robustness against the outlier nodes. Now, we will introduce the system model, assumptions, and parameters underlying DBSCAN with respect to our B-D2D system scenario.

\subsection{System Model and Assumptions}

\begin{figure}
\centering
			\includegraphics[width=1\columnwidth]{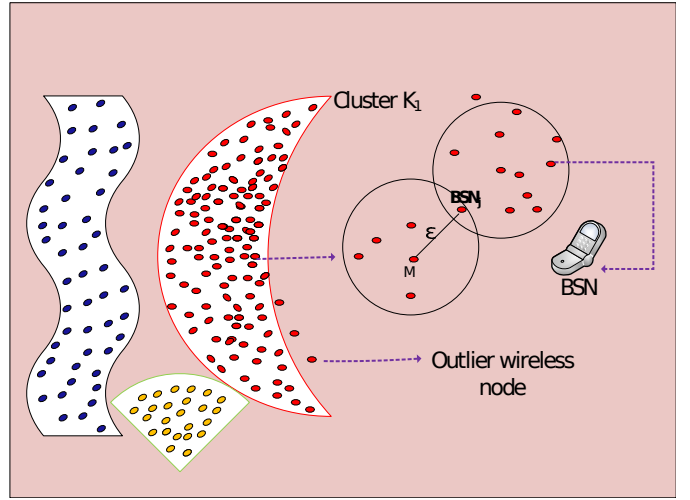}
      \caption{System model using DBSCAN algorithm}\vspace{-4mm}
      \label{fig:systemmodel}
\end{figure}

Let us assume a cellular network, that consists of a $\eta$ number of $BSNs$ and $\omega$ number of base stations  distributed across the geographic cellular space according to a Poisson distribution with densities $\lambda_{\eta}$ and $\lambda_{\omega}$ respectively.  
let $\{K_{1}....K_{\kappa}\}$ be a cluster set containing $\kappa$ clusters, each with different densities, where $\kappa$<<$\eta$. For simplicity and without loss of generality, we consider a cellular network where a BS serves just one of the clusters $K_{1}$ (see \fref{fig:systemmodel}).

\textbf{Parameters:} As we mentioned earlier DBSCAN algorithm requires the specification of two parameters i.e., $\epsilon$ and $MinBSN$. The parameter $\epsilon$ defines the radius of the neighborhood around a mobile user. In our scenario, $\epsilon$ is the maximum transmission range of mobile users to join the cluster $K_1$, which is determined by the power level of each transmitting device. The parameter $MinBSN$ refers to the minimum number of neighbors required to form a cluster within radius $\epsilon$. Since D2D communication requires at least two devices to establish a D2D system, we assume that every mobile user $M$ needs at least one ${BSN_j}$ in its neighborhood to form a cluster. The file sharing across the clusters is prohibited due to the density-reachable and density-connected property of DBSCAN algorithm. Due to number of pages limitations, for the important definitions and steps of DBSCAN clustering algorithm interested readers may refer to ~\cite{ester1996density}. 




For a randomly chosen mobile user $m \in M$ with maximum radius $\epsilon_{max}$, the probability that $\eta$ number of $BSNs$ exists from the content-requesting node is given as:
\begin{align}
P_{\eta} = \frac{ e^{-\lambda \pi \epsilon^2}}{\textsl{$\eta$!}} {(\lambda \pi \epsilon_{max}^2)^ \textsl{$\eta$}}.
\label{eq:probability_of_unavailability}
\end{align} 
For the sake of tractability, we assume that the size of initial-segment is fixed and normalized, and given by $S_{i}=S_{t}$$\times$$b_{i}$. Where $S_{t}$ is the duration of initial-segment of content $i$ in seconds and $b_{i}$ is the bit-rate in Mbps. To achieve the maximum availability-ratio each $BSN$ will store initial-segments $S_{i}=\{S_1, S_2, S_3, . . S_L\}$ of popular video files with probabilities $Q_i$$\in$$\{Q_1,Q_2,....,Q_{L}\}$ subject to availability of of cache space.

We define the availability of initial-segments for some particular video content $i$ as an event in which requesting mobile user $m$ finds the corresponding $BSN$ within a maximum distance $\epsilon_{max}$. The probability that mobile user can find the initial-segment of content $i$ within its maximum transmission range $\epsilon_{max}$ is given as:
\begin{align}
P^{(i)}_{A}(\epsilon_{max})= 1-P^{(i)}_{U}= 1-e^{- \lambda \pi \epsilon_{max}^2}, 
\end{align}
where $P^{(i)}_{U}$ is the probability of unavailability of initial-segment when the requesting mobile user cannot find the $BSN$ caching the initial-segment of desired video content i.e., when $\eta=0$. 
\subsection{Problem Formulation}
The aim of our work is to maximize the average probability of availability of initial-segments of video contents within the transmission range of requesting devices by exploiting the cache capacity of mobile devices. Now, suppose that $R$ number of  requests are made in cluster $K_{i}$ for an initial-segment of content $i$ from a library $L$ with  request probability $P^{(i)}_{r}$. The average number of initial-segments cached in each $BSN$ is represented by $\sum_{i=1}^{L} Q_{i}$. The caching capacity of each $BSN$ is given by $C$. The objective function in our scenario is thus to maximize the average availability ratio of initial-segments which is defined as the ratio of the sum of the $BSNs$ available within the transmission range of each requesting device and the total requests in a cluster $K_{i}$. Mathematically, we have:
\begin{align}
{\text{maximize}}  \frac{1}{R} \sum_{i=1}^{L} (1-e^{- \lambda \pi \epsilon_{max}^2}) P^{(i)}_{r} Q_{i}  
\label{eq:cdn_gain}
\end{align}
subject to:
\begin{align}
 \sum_{i=1}^{L} Q_{i} \leq C,
\end{align}
\begin{align}
Q_{i} \geq 0,  i=1,2,.....,L,
\end{align}
we can solve the Eq.~\eqref{eq:cdn_gain} by employing a conventional brute force search approach. However, we observe that dealing with Eq.~\eqref{eq:cdn_gain} is computationally intractable, as we grow the size of users' population in a cluster. In order to avoid the complexity of brute force search algorithm, we tried to solve the problem heuristically to maximize the objective function by caching the initial-segments of popular files according to popularity distribution and the storage capacity of a $BSN$. 

\begin{table}[t!]
\caption{Simulation Parameters}
\small{
\centering
\begin{tabular}{l|p{6cm}}
\toprule
Parameters & Value\\
\midrule
$\epsilon_{max}$ & 100m \\
$MinBSN$ & 2 \\
$N_{L}$ & 1000 \\
$\lambda_{\eta}$ & 200,...,2500\\
$\beta$ & 0.6, 0.8, 1 \\
$P^{(i)}_{r} $ & 0.6\\
$S_i$ & 15s,20s,30s,60s \\
\bottomrule
\end{tabular}
\label{tbl:params1}}
\end{table}

\section{Evaluation By Simulation Experiments}
 
In the simulation, we study how the number of requests\footnote{Average number of requests is equal to the number of $BSNs$ in a cluster i.e., $\lambda_{\eta}$.}, popularity distribution, caching strategy, and size of initial-segment affect the overall average availability ratio. We estimated this using Monte Carlo simulation by repeating each operation 500 times. The simulation parameters are given in \tref{tbl:params1}.

\begin{figure*}[ht]
     \centering
		\subfloat[Impact of transmission range on the number of clusters \label{fig:Transmissionrange1000}]
        {\includegraphics[width=0.45\textwidth]{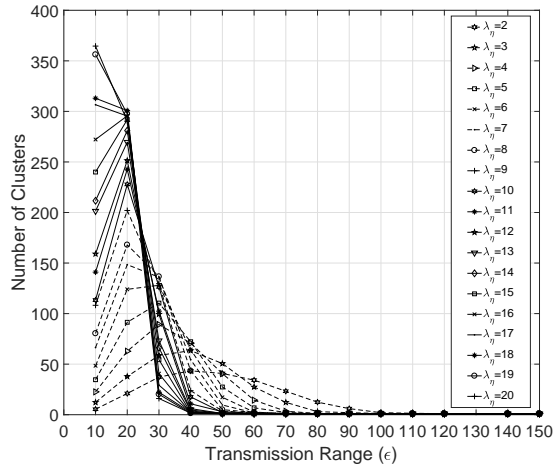}}\hspace{0.5em}
			\subfloat[Impact of transmission range on the number of outlier wireless nodes\label{fig:noise}]
        {\includegraphics[width=0.45\textwidth]{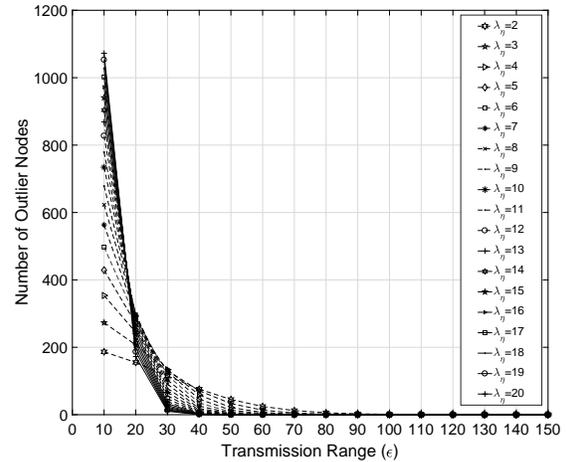}}\hspace{0.5em}
				\caption{Impact of transmission range on the number of clusters and number of outlier wireless nodes}\vspace{-4mm}
		\label{fig:DBSCANSIMULATION}
				\end{figure*}

\subsection{ Simulation configuration}
\begin{figure*}[ht]
     \centering
		\subfloat[The average availability ratio vs number of requests for caching initial-segments and a complete file \label{fig:CCVSCF1}]
        {\includegraphics[width=0.45\textwidth]{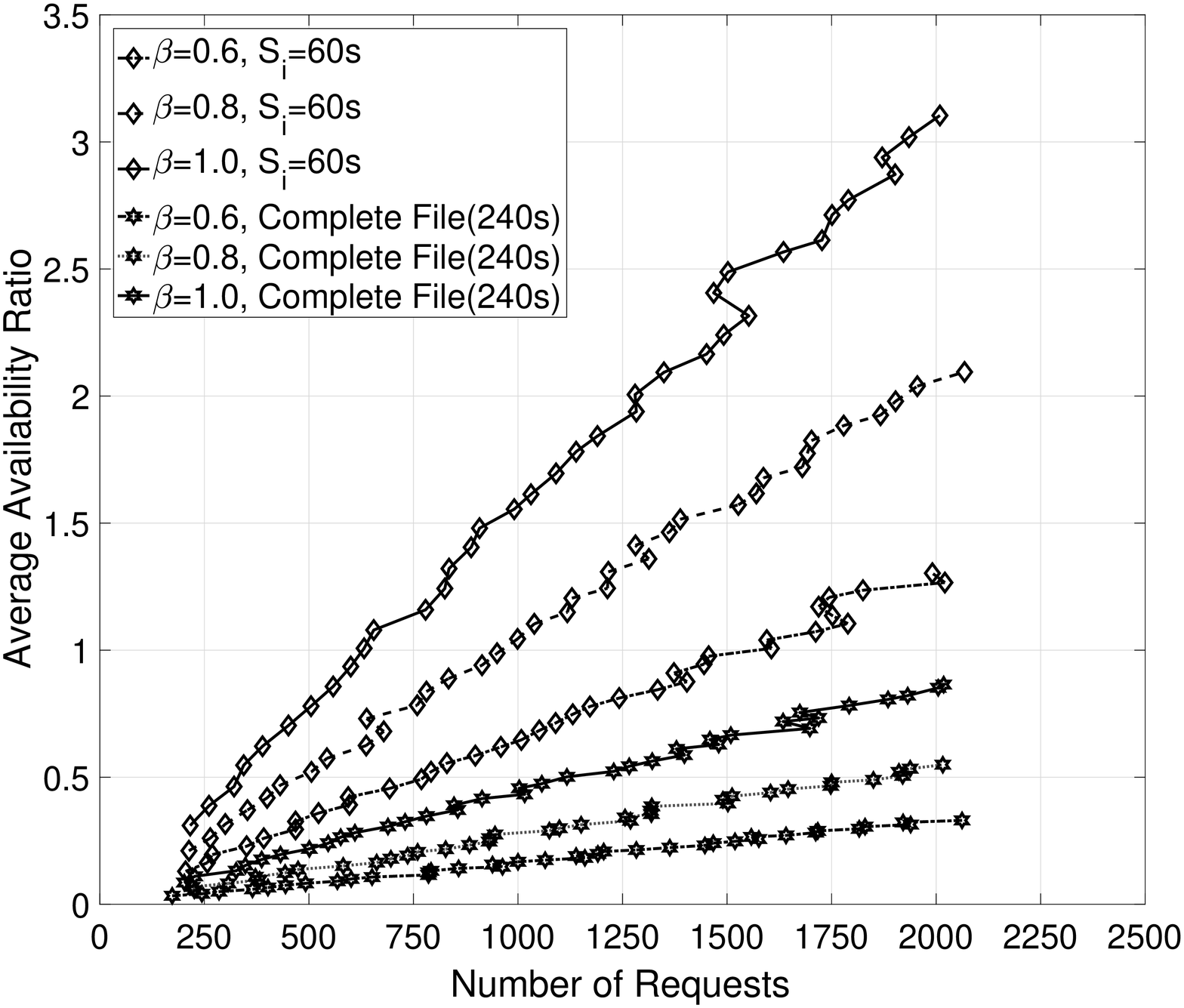}}\hspace{0.5em}
			\subfloat[The average availability ratio vs number of requests for random and MPCO caching \label{fig:random_VS_MPCO}]
        {\includegraphics[width=0.45\textwidth]{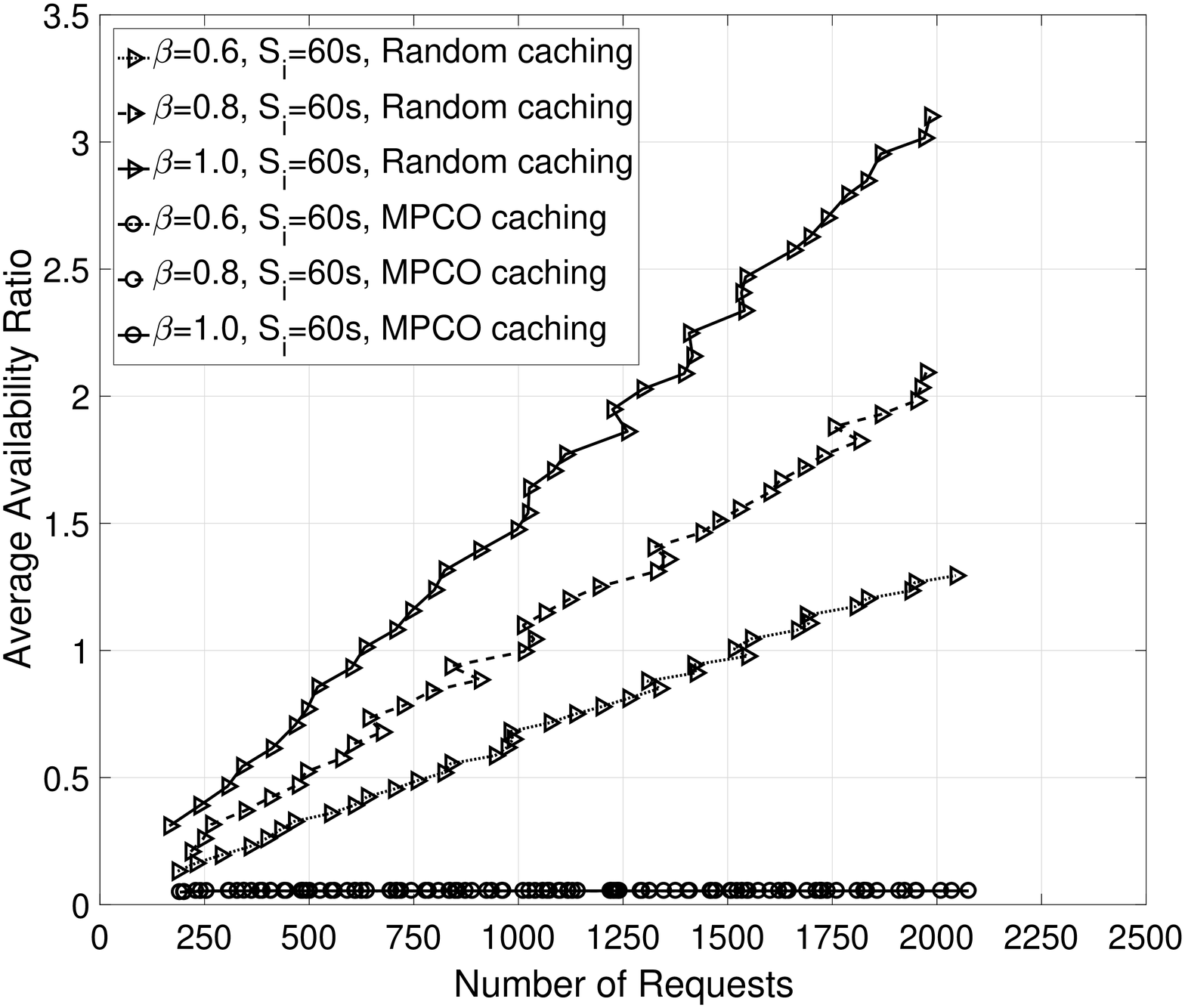}}\hspace{0.5em}
								\subfloat[Impact of size of initial-segments on average availability ratio\label{fig:COMPARISON_OF_CHUNK_CLUSTER_SIZES}]
        {\includegraphics[width=0.45\textwidth]{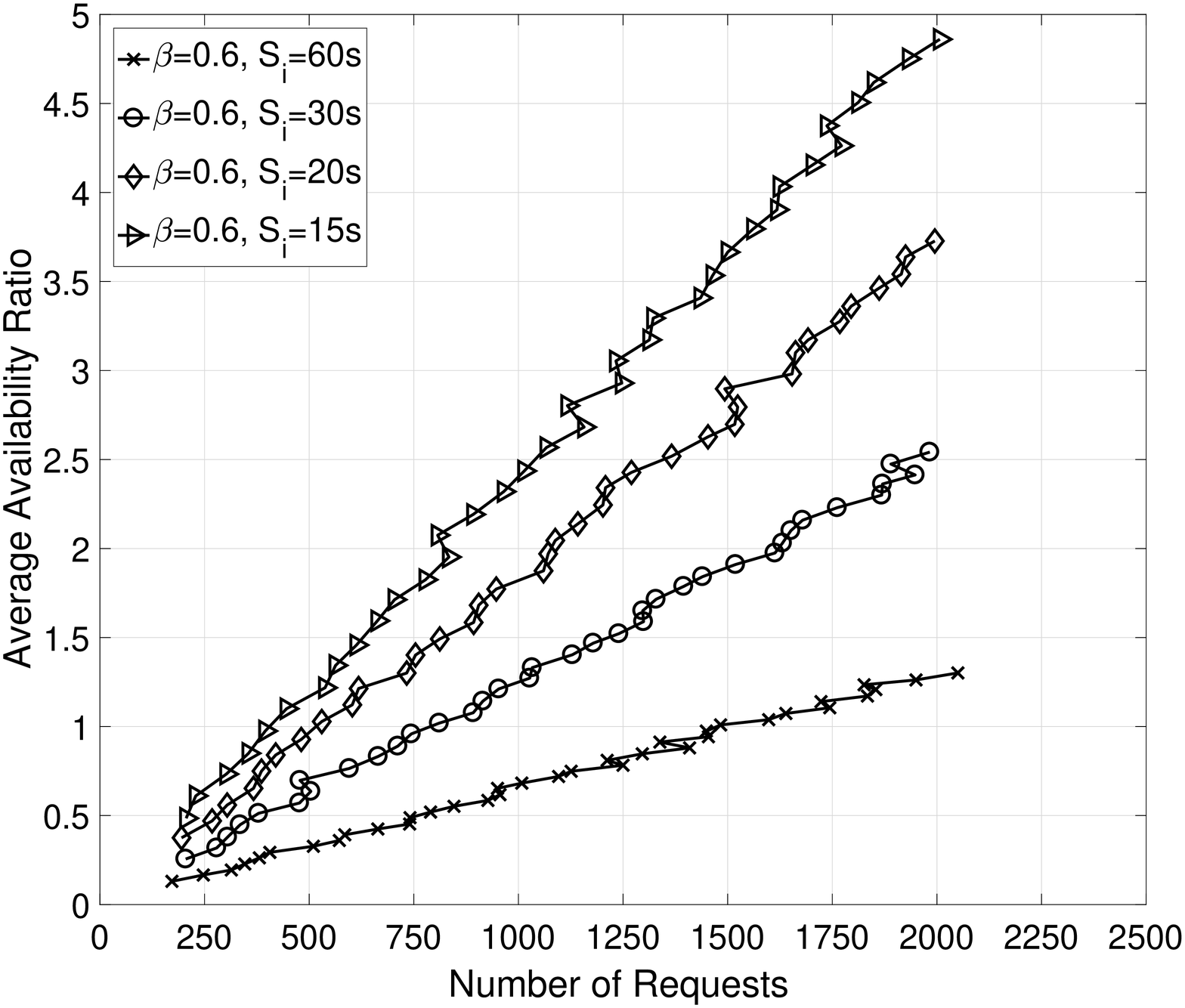}}\hspace{0.5em}
								\subfloat[Impact of size of initial-segment and a complete file on average availability ratio for self-requests\label{fig:self_request1}]
        {\includegraphics[width=0.45\textwidth]{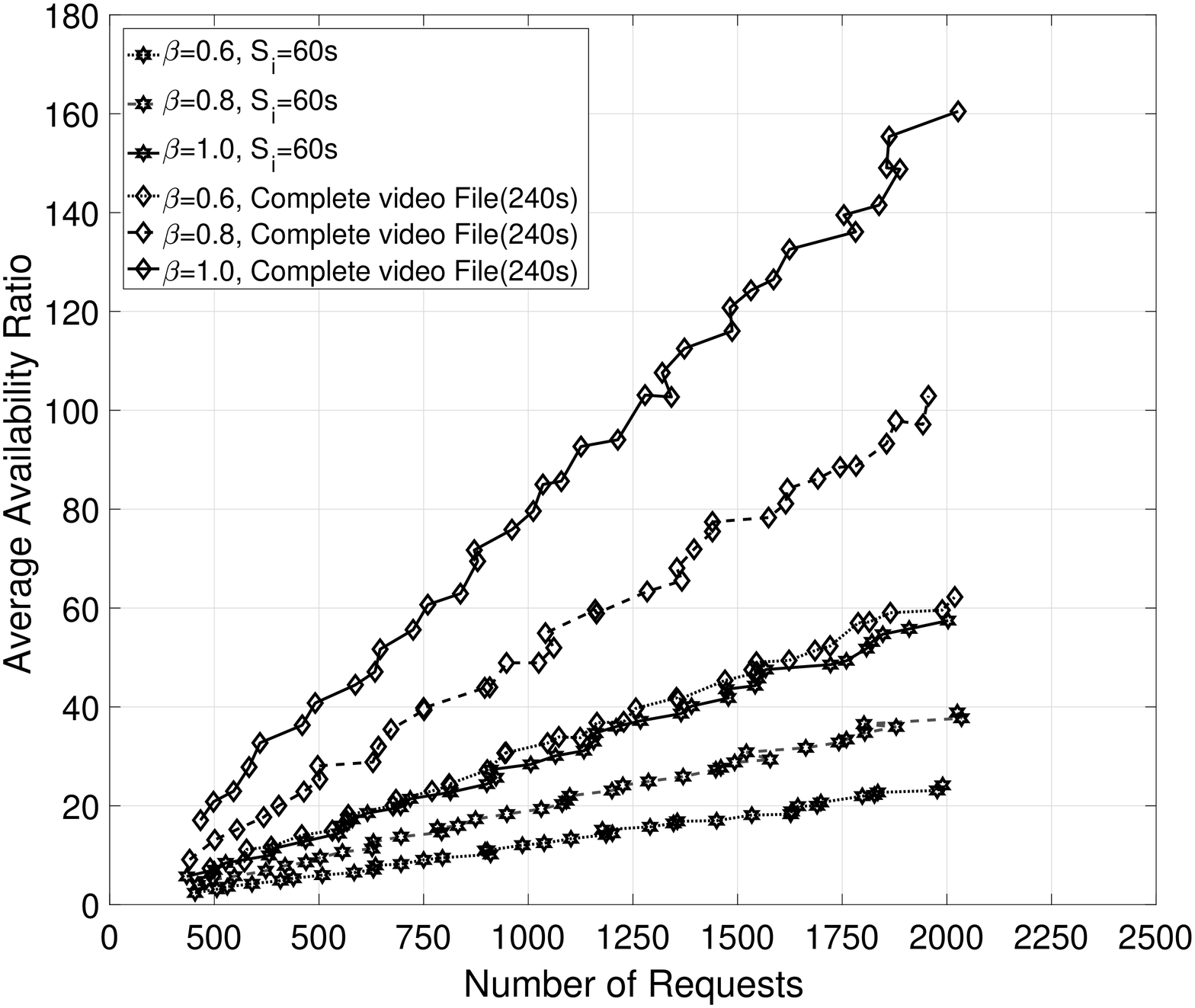}}\hspace{0.5em}
							\caption{Impact of users requests on the average availability ratio for different values of $\beta$, $\lambda_\eta$, size of initial-segments and caching techniques}\vspace{-4mm}
		\label{fig:requestsvsars}
  \end{figure*}
In the simulation, first we distribute the $BSNs$ for different values of $\lambda_{\eta}$=[200,..,2500] in a square area of 1000m x 1000m,  which is a spectrum range of typical urban macro cell. We assign files to graphically distributed $BSNs$ according to the random and MPCO (most-popular-caching-only) caching policy using popularity distribution $\beta$=[0.6,0.8,1] from a library of 1000 distinct video files, and generate file requests at random according to the Zipf distribution $P^{(i)}_{r}$=0.6~\cite{golrezaei2014base}. For comparison, the average availability ratio is used as the performance criterion. According to one study, the average length of top ten YouTube videos is 4 minute approximately~\cite{comScore}. We assume that each $BSN$ can store a single YouTube content and size of initial-segment, .i.e., $S_i$ should not exceed than the average length of YouTube video. Then, we use the DBSCAN algorithm to group $BSNs$ into a single, large and fully connected cluster. 



\subsubsection{Simulation Results} 
DBSCAN algorithm suffers from two fundamental issues: 
\begin{enumerate}
\item Its efficiency and performance entirely depends on the input values of its two parameters .i.e., $\epsilon$ and $MinBSN$. The slight change in their input values may lead to a different set of clusters of a date set.
\item It fails when there are non-uniform density and connected clusters, usually resulted from improper use of $\epsilon$.
\end{enumerate}
Therefore, in order to see the impact of average availability ratio over total requests, first we need to decide the optimum value of $\epsilon$. We conducted experiments in MATLAB and generated two plots which helped us in deciding optimum values of $\epsilon$ that results in minimum or no outlier wireless nodes along with fully connected single large cluster. We set the value of $MinBSN$ to 2, as D2D communication requires at least two devices to establish a D2D network. \fref{fig:DBSCANSIMULATION} shows the impact of transmission range '$epsilon$' on the number of clusters and outlier wireless nodes respectively. 



\fref{fig:Transmissionrange1000} shows the impact of the transmission range, i.e., $\epsilon$ on the number of clusters for different values of $\lambda_\eta$. We can observe from the figure, from transmission range 30 onwards, there is a decrease in the number of clusters. Which means for small value of $\epsilon$ a large number of wireless nodes would subdivide into different groups of clusters. While on the contrary, as the value of transmission range is increased, the density also starts increasing, and more and more wireless nodes are becoming neighbors till the transmission range reaches 100 and onward. So at transmission range 100 and onward, there exists a single and large fully connected cluster.

Now the next thing is to find the number of outlier nodes exists at transmission range 100 and onward. The impact of outliers wireless nodes on transmission range has been examined in ~\fref{fig:noise}. As per ~\fref{fig:noise}, from transmission range 10 onwards there is a decrease in the number of outlier wireless nodes. Which means, as the value of transmission range increases more and more wireless nodes can be assigned to a group of clusters. At transmission range 70 and onward number of outlier wireless nodes are almost negligible. Therefore, based on the experiments the first $\epsilon$ value can be chosen as 100, as it generates a fully connected single large cluster with minimum outlier wireless elements. 

Now, as we have got the optimum value of $\epsilon$, we can run our simulation to get further desired results. \fref{fig:requestsvsars} show the average number of requests versus the average availability ratio for different values of $\beta$, size  of an initial-segment, self-requests and caching techniques. \fref{fig:CCVSCF1} compares the average number of requests versus the average availability ratio for random caching policy when, i) each $BSN$ is caching initial-segments of four popular video files and, ii) a complete YouTube video file. The length of initial-segments of one video file is 60 seconds in this experiment. It is clear from the figure that the average availability ratio increase significantly, when each $BSN$ is caching the initial-segments of four different video files in comparison to caching a complete video file. This figure also confirms that the ratio of average availability for the random caching grows as $\lambda_{\eta}$ and $\beta$ increases. This is because the skewed distribution of users' interest toward small fraction of top ranked content. Moreover, when the number of users grows, requesting device surrounds by more and more $BSNs$ and the probability of finding the initial-segments of desired video content within the $\epsilon$ also become higher. 

\fref{fig:random_VS_MPCO} compares the average number of requests versus the average availability ratio for random and MPCO caching, when each $BSN$ is caching initial-segments of four different video files. This figure confirms that random caching outperforms MPCO scheme. The reason is that in MPCO scheme each $BSN$ stores the initial-segments of top listed popular files redundantly. As a result the chances of finding initial-segments of desire random files will decrease. Although, in random caching the cached initial-segments may overlap, but due to the randomness and large pool of cache items, the average availability ratio for random caching is larger than that for MPCO caching. This figure also depicts the same effects on the average availability ratio for different values of $\lambda_{\eta}$ and $\beta$ as we have described for \fref{fig:CCVSCF1}. Since  random caching shows better performance than MPCO policy, therefore, we will consider the remaining simulation experiments using random caching only.

\fref{fig:COMPARISON_OF_CHUNK_CLUSTER_SIZES} shows how the size of initial-segments, number of requests, and different values of $\beta$ effect on average availability ratio. We performed our experiments for different lengths of initial-segments ranges from 15sec to 60sec. It is noticeable that average availability ratio increases as the size of initial-segment decreases. Which means that, since $BSN$ has cached initial-segments of variety of video files, users get the opportunity to request from a large pool of cached initial-segments and can find and download the initial-segment of desired video content within its vicinity locally in a cluster. \fref{fig:self_request1} shows the impact of self-requests when a mobile user can find and play the requested video file with zero startup delay. As expected, it is obvious from the figure that, the average number of self-requests is higher for B-D2D approach due to presence of initial-segments of multiple video contents in a cache of each $BSN$. So B-D2D approach also has the potential to improve the average self-request ratio when each user cache the video content partially.   

\section{Conclusion}

In this paper, we propose a concept of B-D2D system in which wireless nodes of cellular network can cache the initial-segments of the popular video files and upon request may share to users within their proximity. Since B-D2D caching concept allows partial caching of more video contents, mobile users can discover and download the initial-segments of desired video files from nearby devices with  higher data rates. 

According to one study, the audience of the short duration videos is less tolerate to startup delays comparable to the audience of long duration content~\cite{krishnan2013video}. Moreover, the time span of videos gets shorter because user attentions are decreasing ~\cite{Shortlengthstatistics}, and a significant portion of the wireless cellular spectrum is consumed by short length video streams, such as advertisements, trailers, and users generated content on OSNs. So the B-D2D approach may bring advantages for the audience of short duration videos who are less tolerate to startup delay. But many factors are still need to be investigated as long as achieving low startup delay is our concern, such as finding optimal initial-segment size that results in highest cache-hits and minimum startup delay. In particular, as our future work  we are investigating the impact of our B-D2D approach on startup delay by taking into account the real wireless network conditions such as interference, shadowing and channel fading~\cite{zarringhalam2009jointly}~\cite{olfat2005optimum} etc. 


\bibliographystyle{abbrv}
\bibliography{biblio}
\vspace{2mm}
\bibliographystyle{IEEEtran}

\end{document}